# The model of dark galactic halo based on equilibrium distribution function

Valentin D. Gladush

*Theoretical Physics Department, Dnepropetrovsk National University,*
*Gagarin Ave. 72, Dnepropetrovsk 49010. Ukraine*

**Abstract.** The model of galactic halo as a system of the collisionless small neutral particles from the standpoint of kinetic theory is considered. For the equilibrium system with the use of Maxwell-Boltzmann distribution we obtained results which explain the observed flat rotation curve. These results practically coincide with the results of equilibrium spherical gas cloud model in Newtonian gravitation with linear state equation.

**1 Introduction**

Rotational speed $V$ of the object on a stable Keplerian orbit with radius $r$ around the galaxy can be found from the formula

$$V = \sqrt{\frac{GM}{r}}, \qquad (1)$$

where G – is the gravitational constant. Thus, if $r$ lies outside the visible part of the galaxy, one might expect for a rotation speed to be

$$V \propto \frac{1}{\sqrt{r}}.$$

This disagrees with the data of observational astronomy, where we have $V = const$. For our solar system, this velocity is $V \approx 220 \, km/s$. The above inconsistency is explained by the presence of a hidden mass - dark matter (DM), which is distributed in the form of a dark galactic halo (see e.g. [1,2]). It cannot be otherwise detected, but with its gravitational impact on stars. Let us assume that DM is distributed in the galaxy in spherically symmetric manner. Then the mass of DM in a ball of radius $r$ is $M = M(r)$. Here, $r$ is the distance from the galactic center. Thus, ignoring the visible matter, we have

$$V = \sqrt{\frac{GM(r)}{r}} = const. \qquad (2)$$

Hence

$$M(r) = \frac{V^2}{G} r. \qquad (3)$$

On the other hand, the mass of the DM ball is

$$M(r) = 4\pi \int_{r_0}^{r} r^2 \rho_{DM}(r) dr, \qquad (4)$$

where $\rho_{DM}(r)$ – is the density of the DM, $r_0$ – lower limit of the plateau in the rotation curve. From these two relations it implies following formula for the density of the DM

$$\rho_{DM}(r) = \frac{V^2}{4\pi G r^2}, \quad r > r_0. \qquad (5)$$

Note that the mass per unit of DM is constant in a thin spherical layer of cloud

$$\frac{dM(r)}{dr} = 4\pi r^2 \rho_{DM}(r) = \frac{V^2}{G} = const, \quad r > r_0. \qquad (6)$$

Let us find the gravitational field $\varphi$ of DM. In the Newtonian gravity it is described by the Poisson equation

$$\Delta \varphi = 4\pi G \rho_{DM}. \qquad (7)$$

In the spherically symmetric case, for the density of DM (5) we have

$$\varphi = V^2 \ln r - \frac{C_1}{r} + C_2.$$

The second term of this formula corresponds to the central source. We are looking for the gravitational potential generated by a cloud of DM with $r > r_0$. Therefore, we can set $C_1 = 0$, and choose the constant $C_2$ so that $\varphi(r_0) = 0$. Thus, the assumption of a spherically-symmetric of DM and fact of the plateau on the rotation curves lead to the Newtonian potential

$$\varphi = V^2 \ln \frac{r}{r_0}, \quad r > r_0. \tag{8}$$

Let us now consider the equilibrium conditions of DM. To do this we must enter an assumption about the nature of DM. The simplest assumption is reduced to the introduction of a DM cloud in the form of an ideal gas of nonrelativistic particles with a density $\rho$ and pressure $P$. Then the equilibrium condition for the cloud has the form

$$\nabla P = -\rho_{DM} \nabla \varphi.$$

Hence the equilibrium condition of a spherical cloud of DM follows

$$\frac{dP}{dr} = -\rho_{DM} \frac{d\varphi}{dr}. \tag{9}$$

Using the expression for the density (5), we find the pressure of DM

$$P = \frac{V^4}{8\pi G r^2}, \quad r > r_0. \tag{10}$$

Whence, using (5) we obtain a linear equation for the state of the DM in form

$$P = \frac{1}{2} V^2 \rho_{DM}. \tag{11}$$

The upper boundary of DM cloud can be estimated with the condition $\rho_{DM} \geq \rho_{GDM}$, where $\rho_{GDM}$ is the density of the intergalactic DM. Then, we obtain the radius of the DM cloud

$$r_G = \frac{V}{2\sqrt{\pi G \rho_{GDM}}}$$

It is obvious that the described phenomenological model of TM is not complete. The meanings of DM pressure and temperature are not clear as well.

## 2 Kinetic approach

We consider now a DM collision-free gas of nonrelativistic particles. By assumption, they are the neutral, spinless, massive particles at very small sizes and they can communicate only by gravitation. The scattering cross section of these particles is so small, and the mean free path is so large that they are free to fly through the planets and stars without any undergoing changes. Elementary black holes (BH) with mass of order of the Planck mass can be considered as a candidate of such particles. They can possibly be the remnants of black holes evaporation. Stable elementary black holes may play role of maximally heavy elementary particles and, possibly, DM particles (maximons, friedmon, etc.) [3, 4]. Elementary black holes are characterized by an extremely small scattering cross-section of the order of $10^{-66}$ $cm^2$ [3].

We apply the kinetic approach [2,5] to the aggregate of such particles. Here, the basic value is the distribution function $\psi(\vec{r},\vec{v},t)$, where $\vec{v}$ - is the velocity, $\vec{r}$ – is the radius vector of particles. The number density of particles in space with coordinates $\{\vec{r},\vec{v}\}$ is as follows $f(\vec{r},\vec{v}) = N\psi(\vec{r},\vec{v},t)$, where $N$ - is the total number of particles. The DM mass density is

$$\rho_{DM}(\vec{r},t) = m\int f(\vec{r},\vec{v},t) d\vec{v}, \tag{12}$$

where $m$ - is the mass of the particle. The distribution function satisfies the collisionless kinetic equation

$$\frac{\partial \psi}{\partial t} + \left(\vec{v} \cdot \frac{\partial \psi}{\partial \vec{r}}\right) + \left(\vec{F} \cdot \frac{\partial \psi}{\partial \vec{v}}\right) = 0. \tag{13}$$

Here $\vec{F} = -\nabla \varphi$ – is the gravitational force, $\varphi$ - is the gravitational potential that satisfies the Poisson equation (7). This set of equations is the complete set of Jeans equations [6] which describes self-gravitating collisionless system of particles. If the distribution function $\psi(\vec{r},\vec{v},t)$ is found, then the components of "stress" tensor can be obtained using the formula

$$T_{\alpha\beta} = m\int f\, v_\alpha v_\beta d\vec{v}. \tag{14}$$

In the case of the equilibrium configurations, the distribution function and the mass density are independent of time $\psi = \psi_0(\vec{r},\vec{v})$, $\rho = \rho_0(\vec{r})$, and the equation for the equilibrium distribution function has the form

$$\left(\vec{v}\cdot\frac{\partial\psi_0}{\partial\vec{r}}\right) - \left(\frac{\partial\varphi_0}{\partial\vec{r}}\cdot\frac{\partial\psi_0}{\partial\vec{v}}\right) = 0. \tag{15}$$

where $\varphi_0$ is the self-consistent potential satisfying Poisson's equation $\Delta\varphi_0 = 4\pi G\rho_0$. In this case we deal with anisotropic pressure that is given by formula

$$P_r = m\int f_0 v_r^2 d\vec{v}, \qquad P_t = T_{\theta\theta} = T_{\alpha\alpha} = \frac{1}{2}m\int f_0 v_\perp^2 d\vec{v}. \tag{16}$$

Strictly speaking, this is not the pressure, but the velocity dispersion!

The equilibrium distribution function is a function of energy E and the other possible single-valued integrals of motion. In the isotropic case the distribution function can depend only on the energy $\psi = \psi_0(E)$, in this case $P_r = P_t = P_0$. Then the mass density and pressure are given by following formulae

$$\rho_0 = 4\pi\sqrt{2}m\int_{\varphi_0}^\infty f_0(E)(E-\varphi_0)^{1/2}dE, \qquad P_0 = \frac{8\pi\sqrt{2}}{3}m\int_{\varphi_0}^\infty f_0(E)(E-\varphi_0)^{3/2}dE. \tag{17}$$

Differentiating the last equality in (16) and comparing the result with the previous formula in (16), we get the known condition of hydrodynamic equilibrium

$$\frac{dP_0}{dr} = -\rho_0\frac{d\varphi_0}{dr}. \tag{18}$$

which was used above, when we have considered the equilibrium conditions of a DM cloud. It is the hydrodynamic analogy.

Let us consider a partial solution of the kinetic equation − Maxwell-Boltzmann distribution [7]

$$\psi(\vec{r},\vec{v}) = Ae^{-E/\theta} = \frac{1}{J}\left(\frac{m}{2\pi\theta}\right)^{3/2}\exp\left(-\frac{m}{\theta}\left(\frac{\vec{v}^2}{2}+\varphi_0(\vec{r})\right)\right), \tag{19}$$

where

$$J = \int \exp\left(-\frac{m}{\theta}\varphi_0(\vec{r})\right)d\vec{r}, \tag{20}$$

$\theta = kT$ − is the module of the canonical distribution. The function $\psi(\vec{r},\vec{v})$ - is the probability density of a certain state of the particle. The mean number density of particles in space with coordinates $\{r,v\}$ is described by formula $f(\vec{r},\vec{v}) = N\psi(\vec{r},\vec{v})$, where $N$ - is the total number of particles in the system, so

$$\int f(\vec{r},\vec{v})d\vec{v}d\vec{r} = V.$$

Mass density of the DM in the cloud is given by

$$\rho_{DM}(\vec{r}) = m\int f(\vec{r},\vec{v})d\vec{v} = \frac{mN}{J}\left(\frac{m}{2\pi\theta}\right)^{3/2}\int \exp\left(-\frac{m}{\theta}\left(\frac{\vec{v}^2}{2}+\varphi_0(\vec{r})\right)\right)d\vec{v}, \tag{21}$$

where $m$ - is the mass of DM particle. Then the total mass of the cloud is $M = mN$. Hence we get

$$\rho_{DM}(\vec{r}) = \frac{M}{J}e^{-m\varphi_0/\theta}. \tag{22}$$

The Poisson equation for the self-consistent spherically symmetric field $\varphi_0$ takes the form

$$\Delta\varphi = \frac{d^2\varphi_0}{dr^2} + \frac{2}{r}\frac{d\varphi_0}{dr} = 4\pi G\frac{M}{J}e^{-m\varphi_0/\theta}. \tag{23}$$

This nonlinear equation has the following particular solution

$$\varphi_0 = \varphi_{DM} = \frac{2\theta}{m}\ln\left(\frac{r}{r_0}\right). \tag{24}$$

Herewith the module of the canonical distribution should be equal

$$\theta = 2\pi m G r_0^2 \frac{M}{J}. \qquad (25)$$

Potential (24) is analogous to the potential (8) of the phenomenological model. Using the hydrodynamic analogy and comparing these potentials, we arrive at the formula for the distribution module

$$\theta = mV^2/2. \qquad (26)$$

The above expressions for the modulus of $\theta$, formulae (25) and (26) lead to the relation for the total mass of DM cloud

$$M_{DM} = \frac{JV^2}{4\pi G r_0^2}. \qquad (27)$$

Hence, using (22), (24) for the density of dark matter, we obtain expression (5), which corresponds to the phenomenological picture.

### 3 Conclusions

Thus, as a model of the galactic halo we can consider the collision-free system of very small, very heavy neutral spinless DM particles, which interact only by gravitation. DM particles are subject of the Maxwell-Boltzmann distribution. Galaxy has the atmosphere of DM, i.e. the DM galactic halo, which is actually transparent. The number density of DM particles in the atmosphere is

$$n_{DM}(r) = \frac{M}{mJ} e^{-m\varphi_0/\theta} = \frac{V^2}{4\pi m G r^2}. \qquad (28)$$

Note that the number of particles in a spherical layer of the atmosphere is constant and does not depend on $r$

$$4\pi n_{DM}(r) r^2 = V^2/mG = const. \qquad (29)$$

In the statistical approach, the role of pressure $P$ (10) is the velocity dispersion (16). This pressure provides stability in the phenomenological picture of the halo. This correspondence is provided by the hydrodynamic analogy, which follows from equations (17) and (18).

This work was supported by the grant of the "Cosmomicrophysics" program of the Physics and Astronomy Division of the National Academy of Sciences of Ukraine